\documentstyle[twoside,fleqn,espcrc2]{article}

\newcommand{\AmS}{{\protect\the\textfont2
  A\kern-.1667em\lower.5ex\hbox{M}\kern-.125emS}}

\title{SO(3) monopoles, vortices and confinement in SU(2) gauge theory}

\author{Tam\'as G.\ Kov\'acs and  E.T.\ Tomboulis
  \address{Department of Physics, University of California, Los Angeles,
           CA 90024, USA} }

\begin{document}

\begin{abstract}

We report on further progress in our programme of understanding confinement
in 3d and 4d SU(2) gauge theory in terms of Z(2) monopoles. A sufficient 
condition for confinement was previously translated into Z(2) monopole
correlation inequalities in a related SO(3) gauge theory. We shall discuss
the physical picture underlying this scenario and present some Monte Carlo
evidence concerning the monopole correlation inequalities.   

\end{abstract}

\maketitle
 
It is an old idea that confinement in Z(N) gauge theories can be understood
in terms of Z(N) vortices (also called fluxons) 
linking with the Wilson loops (see e.g.\ 
\cite{Yoneya}). In three dimensions\footnote{Throughout this paper we shall
consider gauge theories in three dimensions but our arguments can be easily
adopted to the physically more relevant 4d case.} 
a Z(N) vortex is a closed stack
of plaquettes (forming a loop) with a nontrivial Z(N) element on them.
These are excitations very much analogous to Peierls contours in the 2d 
Ising model. 

This vortex condensation mechanism was generalised to SU(N) gauge
theories by Mack and Petkova \cite{Mack}.
Here the magnetic vortices also belong to Z(N), the centre of SU(N) 
but there is a major difference as compared to the Abelian case.
While the Abelian magnetic fluxon necessarily has a thickness
of one lattice spacing, in SU(N) lattice gauge theory, 
due to the continuous nature of 
the group, it can be spread out over several lattice spacings. In fact,
it is exactly the spreading of the vortices that makes it possible for
them to survive at weak coupling. It was emphasized by several authors
that the sufficiently fast (exponential) spreading of the magnetic flux
is an essential feature of any confining SU(N) gauge theory \cite{Mack,Yaffe}.
The general picture emerging from these investigations 
was that it is probably the thick,
spread out vortices interlocking with the Wilson loop that are 
responsible for confinement at weak coupling. Unfortunately it is extremely
hard to trace the dynamics of these fluxons at different length scales
and although they provide a nice intuitive picture, not much progress
was made in obtaining a quantitative understanding of the mechanism.

For the past few years there has been a programme to achieve this goal
and rigorously demonstrate that confinement persists down to arbitrarily
small couplings in SU(2) gauge theory \cite{Tomboulis}. 
In the following we shall summarise the main ideas underlying this 
programme. Here we can present only a rough intuitive picture and for
more technical details the reader is referred to \cite{Tomboulis}
and especially \cite{Kovacs}. The latter is the most detailed presentation
so far, although in the context of the analogous 2d SU(2) principal 
chiral model. 

In the original formulation of Mack and Petkova, thick
vortices were detected by introducing a vortex container,
special boundary conditions along a thick torus linking with the Wilson
loop, to trap the vortex inside the torus. Instead of this we essentially
use the whole lattice as a vortex container. The Wilson loop confinement
criterion can be translated into an exponentially fast spreading of a 
vortex winding around the cubic lattice, 
as a function of the finite lattice size \cite{TY}.
In order to detect thick vortices winding around the lattice we use
an equivalent representation of the SU(2) gauge theory in terms of a
dual Ising model (Z(2) gauge theory in the 4d case) interacting with 
an SU(2)/Z(2) gauge theory. Since the action of the SU(2)/Z(2) system 
is insensitive to the Z(2) centre, a thin Z(2) vortex (i.e.\ a stack of
plaquettes) winding around the lattice does not cost an energy proportional
to its length, however due to the periodic boundary conditions 
it signals the presence of a thick vortex winding around the lattice. 
In this way in order to trace thick vortices we have to detect only
thin ones.

Contrary to Z(N) gauge theories, SU(N) gauge theories admit Z(N) monopoles,
elementary lattice cubes that have a net Z(N) flux flowing out of them.
(In the Z(N) case these are absent due to the Bianchi identity.) These
monopoles are gauge invariant objects, not like the U(1) monopoles that
appear in this model only after a partial gauge fixing. For a Monte Carlo 
study  involving both types of monopoles, see \cite{Zsolt}.

The presence of Z(2) monopoles 
means that in our SU(2) model Z(2) vortices can either be closed or bounded
by a pair of monopoles (a monopole loop in 4d). A vortex going all the way
around the lattice is globally quite similar to the one which is bounded
by a nearby pair of monopoles; the difference between them is only a small
local excitation. In the Z(2)$\times$SU(2)/Z(2) formulation the monopoles
of the SU(2)/Z(2) gauge theory turn out to couple to the dual Ising model
and provide an external magnetic field for the Ising spins. 
If they can produce a nonzero effective magnetic field for the dual Ising model
than the Ising spins are ordered which 
in turn means disorder i.e.\ confinement in the
original gauge theory. The presence of a nonzero effective external field
for the dual Ising model can be rigorously established provided that two  
types of monopole correlation inequalities are satisfied. 

At this point we want to emphasize that the presence of 
Z(2) monopoles is not necessary
for confinement. It is of course not small monopoles but large spread out
vortices that are really responsible for confinement. 
However, as we have already seen, some of the large
vortices naturally occur ``tagged'' with a pair of monopoles, i.e.\ they
are not closed but have a small gap instead. It is exactly these tagged 
vortices that couple to the dual Z(2) system and make it possible to 
establish confinement in terms of the Z(2) variables. In other words,
if we were to constrain out Z(2) monopoles, confinement would not be
lost but we would not be able to see it in terms of the Z(2) variables.

Now returning to the monopole inequalities, the first of these is a 
factorisation inequality pertaining
to monopole pairs (loops in 4d), this was discussed to some length in 
\cite{Tomboulis}. The second inequality that we would need to prove
is that the probability of having a nearby monopole pair with their flux
winding around the lattice, is greater than some finite non-zero number.
In other words, the free energy of the above configuration is a bounded
function of the lattice size. A simple semiclassical estimate supports
this assumption in 3 and 4 dimensions (but not above 4). In the remainder
of the present paper we briefly present the results of a Monte Carlo 
measurement of this quantity as a function of the lattice size.

We could do the measurement only in the 3d case since in 4d due to the long
lifetime of the large vortices we could not obtain enough statistics.
We measured the probability of the configuration with two fixed adjecent 
monopoles, their flux winding around the lattice; normalised by the
probability of having no monopoles at all and no vortex going around the
lattice in the given direction. The measurement was done by generating
a series of configurations using a local heat bath algorithm with the
Villain form of the SU(2)/Z(2) action \cite{Villain},
\begin{equation}
 S[U,\sigma]=  \beta \sum_p \sigma_p \mbox{tr} U_p,
\end{equation}  
where the $\sigma_p$'s are Z(2) valued plaquette variables that are 
summed over in the partition function to ensure the desired Z(2) invariance.
\begin{figure}
\setlength{\unitlength}{0.240900pt}
\ifx\plotpoint\undefined\newsavebox{\plotpoint}\fi
\sbox{\plotpoint}{\rule[-0.200pt]{0.400pt}{0.400pt}}%
\begin{picture}(930,629)(20,20)
\font\gnuplot=cmr10 at 10pt
\gnuplot
\sbox{\plotpoint}{\rule[-0.200pt]{0.400pt}{0.400pt}}%
\put(176.0,113.0){\rule[-0.200pt]{0.400pt}{118.764pt}}
\put(176.0,113.0){\rule[-0.200pt]{4.818pt}{0.400pt}}
\put(154,113){\makebox(0,0)[r]{0.001}}
\put(846.0,113.0){\rule[-0.200pt]{4.818pt}{0.400pt}}
\put(176.0,162.0){\rule[-0.200pt]{2.409pt}{0.400pt}}
\put(856.0,162.0){\rule[-0.200pt]{2.409pt}{0.400pt}}
\put(176.0,191.0){\rule[-0.200pt]{2.409pt}{0.400pt}}
\put(856.0,191.0){\rule[-0.200pt]{2.409pt}{0.400pt}}
\put(176.0,212.0){\rule[-0.200pt]{2.409pt}{0.400pt}}
\put(856.0,212.0){\rule[-0.200pt]{2.409pt}{0.400pt}}
\put(176.0,228.0){\rule[-0.200pt]{2.409pt}{0.400pt}}
\put(856.0,228.0){\rule[-0.200pt]{2.409pt}{0.400pt}}
\put(176.0,241.0){\rule[-0.200pt]{2.409pt}{0.400pt}}
\put(856.0,241.0){\rule[-0.200pt]{2.409pt}{0.400pt}}
\put(176.0,252.0){\rule[-0.200pt]{2.409pt}{0.400pt}}
\put(856.0,252.0){\rule[-0.200pt]{2.409pt}{0.400pt}}
\put(176.0,261.0){\rule[-0.200pt]{2.409pt}{0.400pt}}
\put(856.0,261.0){\rule[-0.200pt]{2.409pt}{0.400pt}}
\put(176.0,270.0){\rule[-0.200pt]{2.409pt}{0.400pt}}
\put(856.0,270.0){\rule[-0.200pt]{2.409pt}{0.400pt}}
\put(176.0,277.0){\rule[-0.200pt]{4.818pt}{0.400pt}}
\put(154,277){\makebox(0,0)[r]{0.01}}
\put(846.0,277.0){\rule[-0.200pt]{4.818pt}{0.400pt}}
\put(176.0,327.0){\rule[-0.200pt]{2.409pt}{0.400pt}}
\put(856.0,327.0){\rule[-0.200pt]{2.409pt}{0.400pt}}
\put(176.0,356.0){\rule[-0.200pt]{2.409pt}{0.400pt}}
\put(856.0,356.0){\rule[-0.200pt]{2.409pt}{0.400pt}}
\put(176.0,376.0){\rule[-0.200pt]{2.409pt}{0.400pt}}
\put(856.0,376.0){\rule[-0.200pt]{2.409pt}{0.400pt}}
\put(176.0,392.0){\rule[-0.200pt]{2.409pt}{0.400pt}}
\put(856.0,392.0){\rule[-0.200pt]{2.409pt}{0.400pt}}
\put(176.0,405.0){\rule[-0.200pt]{2.409pt}{0.400pt}}
\put(856.0,405.0){\rule[-0.200pt]{2.409pt}{0.400pt}}
\put(176.0,416.0){\rule[-0.200pt]{2.409pt}{0.400pt}}
\put(856.0,416.0){\rule[-0.200pt]{2.409pt}{0.400pt}}
\put(176.0,426.0){\rule[-0.200pt]{2.409pt}{0.400pt}}
\put(856.0,426.0){\rule[-0.200pt]{2.409pt}{0.400pt}}
\put(176.0,434.0){\rule[-0.200pt]{2.409pt}{0.400pt}}
\put(856.0,434.0){\rule[-0.200pt]{2.409pt}{0.400pt}}
\put(176.0,442.0){\rule[-0.200pt]{4.818pt}{0.400pt}}
\put(154,442){\makebox(0,0)[r]{0.1}}
\put(846.0,442.0){\rule[-0.200pt]{4.818pt}{0.400pt}}
\put(176.0,491.0){\rule[-0.200pt]{2.409pt}{0.400pt}}
\put(856.0,491.0){\rule[-0.200pt]{2.409pt}{0.400pt}}
\put(176.0,520.0){\rule[-0.200pt]{2.409pt}{0.400pt}}
\put(856.0,520.0){\rule[-0.200pt]{2.409pt}{0.400pt}}
\put(176.0,541.0){\rule[-0.200pt]{2.409pt}{0.400pt}}
\put(856.0,541.0){\rule[-0.200pt]{2.409pt}{0.400pt}}
\put(176.0,557.0){\rule[-0.200pt]{2.409pt}{0.400pt}}
\put(856.0,557.0){\rule[-0.200pt]{2.409pt}{0.400pt}}
\put(176.0,570.0){\rule[-0.200pt]{2.409pt}{0.400pt}}
\put(856.0,570.0){\rule[-0.200pt]{2.409pt}{0.400pt}}
\put(176.0,581.0){\rule[-0.200pt]{2.409pt}{0.400pt}}
\put(856.0,581.0){\rule[-0.200pt]{2.409pt}{0.400pt}}
\put(176.0,590.0){\rule[-0.200pt]{2.409pt}{0.400pt}}
\put(856.0,590.0){\rule[-0.200pt]{2.409pt}{0.400pt}}
\put(176.0,598.0){\rule[-0.200pt]{2.409pt}{0.400pt}}
\put(856.0,598.0){\rule[-0.200pt]{2.409pt}{0.400pt}}
\put(176.0,606.0){\rule[-0.200pt]{4.818pt}{0.400pt}}
\put(154,606){\makebox(0,0)[r]{1}}
\put(846.0,606.0){\rule[-0.200pt]{4.818pt}{0.400pt}}
\put(176.0,113.0){\rule[-0.200pt]{0.400pt}{4.818pt}}
\put(176,68){\makebox(0,0){0}}
\put(176.0,586.0){\rule[-0.200pt]{0.400pt}{4.818pt}}
\put(291.0,113.0){\rule[-0.200pt]{0.400pt}{4.818pt}}
\put(291,68){\makebox(0,0){2}}
\put(291.0,586.0){\rule[-0.200pt]{0.400pt}{4.818pt}}
\put(406.0,113.0){\rule[-0.200pt]{0.400pt}{4.818pt}}
\put(406,68){\makebox(0,0){4}}
\put(406.0,586.0){\rule[-0.200pt]{0.400pt}{4.818pt}}
\put(521.0,113.0){\rule[-0.200pt]{0.400pt}{4.818pt}}
\put(521,68){\makebox(0,0){6}}
\put(521.0,586.0){\rule[-0.200pt]{0.400pt}{4.818pt}}
\put(636.0,113.0){\rule[-0.200pt]{0.400pt}{4.818pt}}
\put(636,68){\makebox(0,0){8}}
\put(636.0,586.0){\rule[-0.200pt]{0.400pt}{4.818pt}}
\put(751.0,113.0){\rule[-0.200pt]{0.400pt}{4.818pt}}
\put(751,68){\makebox(0,0){10}}
\put(751.0,586.0){\rule[-0.200pt]{0.400pt}{4.818pt}}
\put(866.0,113.0){\rule[-0.200pt]{0.400pt}{4.818pt}}
\put(866,68){\makebox(0,0){12}}
\put(866.0,586.0){\rule[-0.200pt]{0.400pt}{4.818pt}}
\put(176.0,113.0){\rule[-0.200pt]{166.221pt}{0.400pt}}
\put(866.0,113.0){\rule[-0.200pt]{0.400pt}{118.764pt}}
\put(176.0,606.0){\rule[-0.200pt]{166.221pt}{0.400pt}}
\put(521,23){\makebox(20,-10){\large $\beta$}}
\put(176.0,113.0){\rule[-0.200pt]{0.400pt}{118.764pt}}
\put(234,554){\raisebox{-.8pt}{\makebox(0,0){$\Diamond$}}}
\put(245,554){\raisebox{-.8pt}{\makebox(0,0){$\Diamond$}}}
\put(257,553){\raisebox{-.8pt}{\makebox(0,0){$\Diamond$}}}
\put(268,553){\raisebox{-.8pt}{\makebox(0,0){$\Diamond$}}}
\put(280,552){\raisebox{-.8pt}{\makebox(0,0){$\Diamond$}}}
\put(291,551){\raisebox{-.8pt}{\makebox(0,0){$\Diamond$}}}
\put(303,551){\raisebox{-.8pt}{\makebox(0,0){$\Diamond$}}}
\put(314,550){\raisebox{-.8pt}{\makebox(0,0){$\Diamond$}}}
\put(326,549){\raisebox{-.8pt}{\makebox(0,0){$\Diamond$}}}
\put(337,548){\raisebox{-.8pt}{\makebox(0,0){$\Diamond$}}}
\put(349,547){\raisebox{-.8pt}{\makebox(0,0){$\Diamond$}}}
\put(360,546){\raisebox{-.8pt}{\makebox(0,0){$\Diamond$}}}
\put(372,545){\raisebox{-.8pt}{\makebox(0,0){$\Diamond$}}}
\put(383,543){\raisebox{-.8pt}{\makebox(0,0){$\Diamond$}}}
\put(395,542){\raisebox{-.8pt}{\makebox(0,0){$\Diamond$}}}
\put(406,540){\raisebox{-.8pt}{\makebox(0,0){$\Diamond$}}}
\put(418,538){\raisebox{-.8pt}{\makebox(0,0){$\Diamond$}}}
\put(429,536){\raisebox{-.8pt}{\makebox(0,0){$\Diamond$}}}
\put(441,534){\raisebox{-.8pt}{\makebox(0,0){$\Diamond$}}}
\put(452,532){\raisebox{-.8pt}{\makebox(0,0){$\Diamond$}}}
\put(464,529){\raisebox{-.8pt}{\makebox(0,0){$\Diamond$}}}
\put(475,526){\raisebox{-.8pt}{\makebox(0,0){$\Diamond$}}}
\put(487,523){\raisebox{-.8pt}{\makebox(0,0){$\Diamond$}}}
\put(498,519){\raisebox{-.8pt}{\makebox(0,0){$\Diamond$}}}
\put(510,515){\raisebox{-.8pt}{\makebox(0,0){$\Diamond$}}}
\put(521,511){\raisebox{-.8pt}{\makebox(0,0){$\Diamond$}}}
\put(533,506){\raisebox{-.8pt}{\makebox(0,0){$\Diamond$}}}
\put(544,501){\raisebox{-.8pt}{\makebox(0,0){$\Diamond$}}}
\put(556,495){\raisebox{-.8pt}{\makebox(0,0){$\Diamond$}}}
\put(567,490){\raisebox{-.8pt}{\makebox(0,0){$\Diamond$}}}
\put(579,484){\raisebox{-.8pt}{\makebox(0,0){$\Diamond$}}}
\put(590,477){\raisebox{-.8pt}{\makebox(0,0){$\Diamond$}}}
\put(602,469){\raisebox{-.8pt}{\makebox(0,0){$\Diamond$}}}
\put(613,462){\raisebox{-.8pt}{\makebox(0,0){$\Diamond$}}}
\put(625,454){\raisebox{-.8pt}{\makebox(0,0){$\Diamond$}}}
\put(636,446){\raisebox{-.8pt}{\makebox(0,0){$\Diamond$}}}
\put(648,437){\raisebox{-.8pt}{\makebox(0,0){$\Diamond$}}}
\put(659,428){\raisebox{-.8pt}{\makebox(0,0){$\Diamond$}}}
\put(671,418){\raisebox{-.8pt}{\makebox(0,0){$\Diamond$}}}
\put(682,411){\raisebox{-.8pt}{\makebox(0,0){$\Diamond$}}}
\put(694,399){\raisebox{-.8pt}{\makebox(0,0){$\Diamond$}}}
\put(705,389){\raisebox{-.8pt}{\makebox(0,0){$\Diamond$}}}
\put(717,377){\raisebox{-.8pt}{\makebox(0,0){$\Diamond$}}}
\put(728,366){\raisebox{-.8pt}{\makebox(0,0){$\Diamond$}}}
\put(740,355){\raisebox{-.8pt}{\makebox(0,0){$\Diamond$}}}
\put(751,349){\raisebox{-.8pt}{\makebox(0,0){$\Diamond$}}}
\put(763,334){\raisebox{-.8pt}{\makebox(0,0){$\Diamond$}}}
\put(774,319){\raisebox{-.8pt}{\makebox(0,0){$\Diamond$}}}
\put(786,312){\raisebox{-.8pt}{\makebox(0,0){$\Diamond$}}}
\put(797,299){\raisebox{-.8pt}{\makebox(0,0){$\Diamond$}}}
\put(809,283){\raisebox{-.8pt}{\makebox(0,0){$\Diamond$}}}
\put(820,283){\raisebox{-.8pt}{\makebox(0,0){$\Diamond$}}}
\put(832,269){\raisebox{-.8pt}{\makebox(0,0){$\Diamond$}}}
\put(843,258){\raisebox{-.8pt}{\makebox(0,0){$\Diamond$}}}
\put(855,231){\raisebox{-.8pt}{\makebox(0,0){$\Diamond$}}}
\put(866,214){\raisebox{-.8pt}{\makebox(0,0){$\Diamond$}}}
\end{picture}
\caption{The Z(2) monopole density as a function of $\beta$ on a 
$8^3$ lattice.
\label{fig:mdens}}
\end{figure}
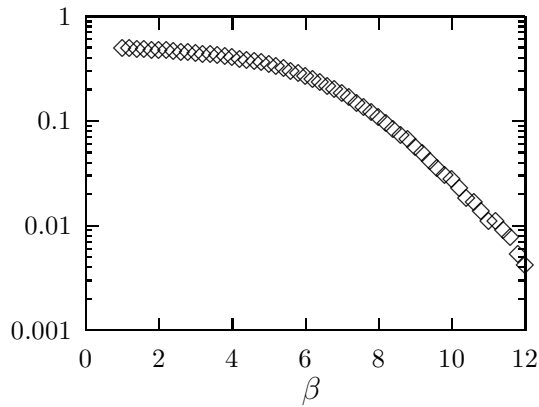
We had to ensure that the $\beta$ we chose was already in the weak coupling
region since we were interested in the weak coupling behaviour of the 
quantity measured. Unfortunately in 3d the crossover between weak and
strong coupling is not so sharp as it is in 4d and in 3d the
specific heat peak at the crossover is missing. Therefore we
used the exponential falling of the monopole density as a criterion 
for weak coupling. We chose $\beta=11.5$, as can be seen in Figure 
\ref{fig:mdens}, this is already well in the exponential regime of 
the monopole density.

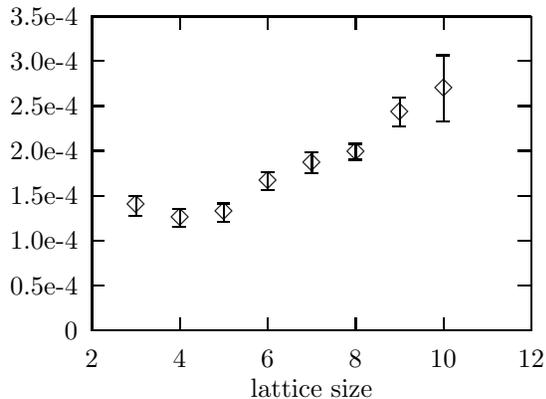
\begin{figure}
\setlength{\unitlength}{0.240900pt}
\ifx\plotpoint\undefined\newsavebox{\plotpoint}\fi
\sbox{\plotpoint}{\rule[-0.200pt]{0.400pt}{0.400pt}}%
\begin{picture}(930,629)(20,20)
\font\gnuplot=cmr10 at 10pt
\gnuplot
\sbox{\plotpoint}{\rule[-0.200pt]{0.400pt}{0.400pt}}%
\put(176.0,113.0){\rule[-0.200pt]{166.221pt}{0.400pt}}
\put(176.0,113.0){\rule[-0.200pt]{4.818pt}{0.400pt}}
\put(154,113){\makebox(0,0)[r]{0}}
\put(846.0,113.0){\rule[-0.200pt]{4.818pt}{0.400pt}}
\put(176.0,183.0){\rule[-0.200pt]{4.818pt}{0.400pt}}
\put(154,183){\makebox(0,0)[r]{0.5e-4}}
\put(846.0,183.0){\rule[-0.200pt]{4.818pt}{0.400pt}}
\put(176.0,254.0){\rule[-0.200pt]{4.818pt}{0.400pt}}
\put(154,254){\makebox(0,0)[r]{1.0e-4}}
\put(846.0,254.0){\rule[-0.200pt]{4.818pt}{0.400pt}}
\put(176.0,324.0){\rule[-0.200pt]{4.818pt}{0.400pt}}
\put(154,324){\makebox(0,0)[r]{1.5e-4}}
\put(846.0,324.0){\rule[-0.200pt]{4.818pt}{0.400pt}}
\put(176.0,395.0){\rule[-0.200pt]{4.818pt}{0.400pt}}
\put(154,395){\makebox(0,0)[r]{2.0e-4}}
\put(846.0,395.0){\rule[-0.200pt]{4.818pt}{0.400pt}}
\put(176.0,465.0){\rule[-0.200pt]{4.818pt}{0.400pt}}
\put(154,465){\makebox(0,0)[r]{2.5e-4}}
\put(846.0,465.0){\rule[-0.200pt]{4.818pt}{0.400pt}}
\put(176.0,536.0){\rule[-0.200pt]{4.818pt}{0.400pt}}
\put(154,536){\makebox(0,0)[r]{3.0e-4}}
\put(846.0,536.0){\rule[-0.200pt]{4.818pt}{0.400pt}}
\put(176.0,606.0){\rule[-0.200pt]{4.818pt}{0.400pt}}
\put(154,606){\makebox(0,0)[r]{3.5e-4}}
\put(846.0,606.0){\rule[-0.200pt]{4.818pt}{0.400pt}}
\put(176.0,113.0){\rule[-0.200pt]{0.400pt}{4.818pt}}
\put(176,68){\makebox(0,0){2}}
\put(176.0,586.0){\rule[-0.200pt]{0.400pt}{4.818pt}}
\put(314.0,113.0){\rule[-0.200pt]{0.400pt}{4.818pt}}
\put(314,68){\makebox(0,0){4}}
\put(314.0,586.0){\rule[-0.200pt]{0.400pt}{4.818pt}}
\put(452.0,113.0){\rule[-0.200pt]{0.400pt}{4.818pt}}
\put(452,68){\makebox(0,0){6}}
\put(452.0,586.0){\rule[-0.200pt]{0.400pt}{4.818pt}}
\put(590.0,113.0){\rule[-0.200pt]{0.400pt}{4.818pt}}
\put(590,68){\makebox(0,0){8}}
\put(590.0,586.0){\rule[-0.200pt]{0.400pt}{4.818pt}}
\put(728.0,113.0){\rule[-0.200pt]{0.400pt}{4.818pt}}
\put(728,68){\makebox(0,0){10}}
\put(728.0,586.0){\rule[-0.200pt]{0.400pt}{4.818pt}}
\put(866.0,113.0){\rule[-0.200pt]{0.400pt}{4.818pt}}
\put(866,68){\makebox(0,0){12}}
\put(866.0,586.0){\rule[-0.200pt]{0.400pt}{4.818pt}}
\put(176.0,113.0){\rule[-0.200pt]{166.221pt}{0.400pt}}
\put(866.0,113.0){\rule[-0.200pt]{0.400pt}{118.764pt}}
\put(176.0,606.0){\rule[-0.200pt]{166.221pt}{0.400pt}}
\put(521,23){\makebox(0,0){lattice size}}
\put(176.0,113.0){\rule[-0.200pt]{0.400pt}{118.764pt}}
\put(245,309){\raisebox{-.8pt}{\makebox(0,0){$\Diamond$}}}
\put(314,289){\raisebox{-.8pt}{\makebox(0,0){$\Diamond$}}}
\put(383,298){\raisebox{-.8pt}{\makebox(0,0){$\Diamond$}}}
\put(452,347){\raisebox{-.8pt}{\makebox(0,0){$\Diamond$}}}
\put(521,376){\raisebox{-.8pt}{\makebox(0,0){$\Diamond$}}}
\put(590,393){\raisebox{-.8pt}{\makebox(0,0){$\Diamond$}}}
\put(659,455){\raisebox{-.8pt}{\makebox(0,0){$\Diamond$}}}
\put(728,493){\raisebox{-.8pt}{\makebox(0,0){$\Diamond$}}}
\put(245.0,293.0){\rule[-0.200pt]{0.400pt}{7.468pt}}
\put(235.0,293.0){\rule[-0.200pt]{4.818pt}{0.400pt}}
\put(235.0,324.0){\rule[-0.200pt]{4.818pt}{0.400pt}}
\put(314.0,275.0){\rule[-0.200pt]{0.400pt}{6.745pt}}
\put(304.0,275.0){\rule[-0.200pt]{4.818pt}{0.400pt}}
\put(304.0,303.0){\rule[-0.200pt]{4.818pt}{0.400pt}}
\put(383.0,283.0){\rule[-0.200pt]{0.400pt}{6.986pt}}
\put(373.0,283.0){\rule[-0.200pt]{4.818pt}{0.400pt}}
\put(373.0,312.0){\rule[-0.200pt]{4.818pt}{0.400pt}}
\put(452.0,333.0){\rule[-0.200pt]{0.400pt}{6.745pt}}
\put(442.0,333.0){\rule[-0.200pt]{4.818pt}{0.400pt}}
\put(442.0,361.0){\rule[-0.200pt]{4.818pt}{0.400pt}}
\put(521.0,360.0){\rule[-0.200pt]{0.400pt}{7.950pt}}
\put(511.0,360.0){\rule[-0.200pt]{4.818pt}{0.400pt}}
\put(511.0,393.0){\rule[-0.200pt]{4.818pt}{0.400pt}}
\put(590.0,381.0){\rule[-0.200pt]{0.400pt}{6.022pt}}
\put(580.0,381.0){\rule[-0.200pt]{4.818pt}{0.400pt}}
\put(580.0,406.0){\rule[-0.200pt]{4.818pt}{0.400pt}}
\put(659.0,433.0){\rule[-0.200pt]{0.400pt}{10.840pt}}
\put(649.0,433.0){\rule[-0.200pt]{4.818pt}{0.400pt}}
\put(649.0,478.0){\rule[-0.200pt]{4.818pt}{0.400pt}}
\put(728.0,441.0){\rule[-0.200pt]{0.400pt}{25.054pt}}
\put(718.0,441.0){\rule[-0.200pt]{4.818pt}{0.400pt}}
\put(718.0,545.0){\rule[-0.200pt]{4.818pt}{0.400pt}}
\end{picture}
\caption{ The probability of two adjecent monopoles with their 
connecting fluxon going around the lattice divided by the probability 
of having no monopoles at all and no fluxon going around the lattice 
in the given direction.
\label{fig:Prob}}
\end{figure}

Since (at a given $\beta$) on larger lattices there are typically more 
monopoles, the probability of the confugurations with exactly one an
zero monopole pairs decreased very rapidly with increasing lattice size.
This meant that the quantity we measured was given as a ratio of two 
numbers, both becoming very small on larger lattices, however their ratio
was expected to be fairly stable. This made the signal less accurate 
on larger lattices and eventually prevented us from going beyond a
lattice size of 10$^3$. Even at this point we typically needed several
thousand configurations to get a signal at all. Our results are
summarised in Figure \ref{fig:Prob} which shows the measured probability as a 
function of the lattice size. As we expect, the probability of
the two-monopole configuration with the long fluxon does not decrease with the
lattice size. This is consistent with the semiclassical estimate 
and makes it quite improbable that it can go to zero as $L \rightarrow
\infty$.

If confinement is present we expect an 
exponential spreading of the flux, i.e. its free energy to approach 
zero exponentially on sufficiently large lattices. In our framework
we would need to prove only the substantially weaker condition that 
the flux free energy does not diverge with the lattice size. It is
remarkable that even this weaker property is absolutely nontrivial
to prove rigorously and we had to resort to Monte Carlo. 

Finally we would like to point out that each step in the prgramme 
outlined here has been given precise formulation and to complete 
it we would need to prove the correlation inequalities discussed above. 

T.G.K.\ thanks the Department of Physics, UCLA and the 
Organisers of Lattice '96 for financial support. He also 
thanks Zsolt Schram for discussions
and help with the Monte Carlo code.


\begin{thebibliography}{9}


\bibitem{Yoneya} Yoneya, T., Nucl.\ Phys.\ {\bf B144} (1978) 195

\bibitem{Mack} Mack, G.\ and Petkova, V.B., Ann.\ Phys.\ {\bf 123} (1979) 442;
Ann.\ Phys.\ {\bf 125} (1980) 117; Z.\ Phys.\ {\bf C12} (1982) 177

\bibitem{Yaffe} Yaffe, L.G., Phys.\ Rev.\ {\bf D21} (1979) 1574

\bibitem{Tomboulis} Tomboulis, E.T., Phys.\ Lett.\ {\bf B303} (1993) 103;
Nucl.\ Phys.\ B Proc.\ Suppl.\ {\bf 30} (1993) 549;
Nucl.\ Phys.\ B Proc.\ Suppl.\ {\bf 34} (1994) 192

\bibitem{TY} Tomboulis, E.T.\ and Yaffe, L.G., Commun.\ Math.\
Phys.\ {\bf 100} (1985) 313 

\bibitem{Zsolt} Daruka, I., Polonyi, J.\ and Schram, Zs., in preparation.

\bibitem{Villain} Caneschi, L., Halliday, I.G. and Schwimmer, A.,
Nucl.\ Phys.\ {\bf B200[FS4]} (1982) 409; Halliday, I.G.\ and Schwimmer,
A., Phys.\ Lett.\ {\bf B102} (1981) 337

\bibitem{Kovacs} Kov\'acs, T.G., SO(3) vortices as a mechanism for generating
a mass gap in the 2d SU(2) principal chiral model, preprint UCLA/96/TEP/12
and hep-lat/9603022 

\end{thebibliography}
\end{document}